\documentclass[twoside,a4paper]{article}
\usepackage{amsmath,graphicx,amssymb,fancyhdr,amsthm,,textcomp,hyperref}
\usepackage{float}
\usepackage{subcaption}
\usepackage[english]{babel}
\usepackage[numbers]{natbib}
\usepackage[acronym, nopostdot,nomain,nonumberlist,numberedsection]{glossaries}

\makenoidxglossaries

\theoremstyle{definition} 
  
\theoremstyle{remark}  
  
\def\beq{\begin{eqnarray}}  
\def\eeq{\end{eqnarray}}  
\def\bsp{\begin{split}}  
\def\esp{\end{split}}

\graphicspath{C:\Users\Deedee\Dropbox\Projects\Blackhole Invariants\ADA Project\KT paper\Figures}
 
\begin{document}  
  
\title{\Large\textbf{Geometric Horizons in the Kastor-Traschen Multi Black Hole Solutions}}  
\author{{\large\textbf{D. D. McNutt$^{1}$  and A. A. Coley$^{2}$ }}
\vspace{0.3cm} \\ 
$^{1}$ Faculty of Science and Technology,\\
University of Stavanger, 
N-4036 Stavanger, Norway  \\
\vspace{0.3cm} \\ 
$^{2}$Department of Mathematics and Statistics,\\
Dalhousie University,
Halifax, Nova Scotia,\\
Canada B3H 3J5
 \vspace{0.3cm} \\
\texttt{david.d.mcnutt@uis.no,aac@mathstat.dal.ca}}  
\date{\today}  
\maketitle  
\pagestyle{fancy}  
\fancyhead{} 
\fancyhead[EC]{}  
\fancyhead[EL,OR]{\thepage}  
\fancyhead[OC]{}  
\fancyfoot{} 
  
\begin{abstract}   
We investigate the existence of invariantly defined quasi-local hypersurfaces in the Kastor-Traschen solution containing $N$ charge-equal-to-mass black holes. These hypersurfaces are characterized by the vanishing of particular curvature invariants, known as Cartan invariants, which are generated using the frame approach. The Cartan invariants of interest describe the expansion of the outgoing and ingoing null vectors belonging to the invariant null frame arising from the Cartan-Karlhede algorithm. We show that the evolution of the hypersurfaces surrounding the black holes depends on an upper-bound on the total mass for the case of two and three equal mass black holes. We discuss the results in the context of the geometric horizon conjectures. 

\end{abstract} 

\section{Introduction}

The event horizon is a defining feature of black hole solutions in General Relativity (GR).  It is defined as the boundary of the non-empty complement of the causal past of future null infinity; i.e., the region for which signals sent from the interior will never escape. The event horizon is typically identified as the surface of the black hole and relates its area to the entropy of the black hole. However, the event horizon is essentially a {\em teleological} object, as we must know the global behaviour of the spacetime in order to determine the event horizon locally \cite{AshtekarKrishnan}. To examine the interaction of realistic black holes with their environment in  numerical GR \cite{T}, in the 3+1 approach  or in the Cauchy-problem in GR, it is necessary to locate a black hole locally \cite{AMS2005,jaramillo2013}. A local characterization may not rely on the existence of an event horizon alone, as black holes are expected to undergo evolutionary processes and are typically dynamical. 

To address this Penrose proposed the concept of {\em closed trapped surfaces without border}, which are compact spacelike surfaces such that the expansions of the future-pointing null normal vectors are negative  \cite{P2}. Consequently, to move from stationary black holes to time-dependent situations, the event horizons (which are Killing horizons, and hence null surfaces) are replaced in practice by {\it apparent horizons} defined as the locus of the vanishing expansion of a null geodesic congruence emanating from a trapped surface $S$ with spherical topology \cite{Booth2005}. A related concept to trapping surfaces are marginally outer (inner) trapped surfaces (MOTSs or MITSs) which are two-dimensional (2D) surfaces for which the expansion $\theta_{(+)}$ ($\theta_{(-)} $) of the outgoing (ingoing) null vector normal to the surfaces vanishes. Assuming a smooth time evolution for the MOTSs (MITSs), the 2D surfaces can be combined to construct a three-dimensional (3D) surface known as a marginally trapped tube (MTT) \cite{booth2007isolated}. If the MTT is foliated by MOTSs for which $\theta_{(-)} <0$ as well, then it is called a dynamical horizon \cite{AshtekarKrishnan}.

Unlike the event horizon, the apparent horizon and MTTs are quasi-local, and they are intrinsically foliation-dependent. In numerical studies of collapse, the teleological nature of  event horizons makes the apparent horizon a more practical surface to track.  Apparent horizons are employed in simulations of high precision waveforms of gravitational waves arising from the merger of compact-object binary systems or in stellar collapse to form black holes in numerical relativity, and the observations by the LIGO collaboration of gravitational waves from black hole mergers relied upon the numerical simulations based on apparent horizons \cite{LIGO}. Due to the foliation dependence of the apparent horizons and MTTs, they are observer dependent, which can lead to ambiguities if care is not taken to relate the differing observers' reference frames. For example, if a MOTS is taken as the outermost trapped surfaces in some foliation of hypersurfaces, $\Sigma_t$, where this foliation is defined as hypersurfaces  of ``constant time'' as determined by a set of observers with clocks that were synchronized on some initial hypersuface, then different sets of observers will observe different MTTs.  For this reason it is important to identify alternative surfaces that are defined invariantly.

For a stationary black hole spacetime, if we know the Killing vector field which acts as the null generator on the event horizon then the horizon is defined locally. This is reflected in the curvature invariants as there is a general procedure to produce scalar polynomial curvature invariants (SPIs) which will vanish on the stationary horizon  \cite{AbdelqaderLake2015,PageShoom2015} or by employing Cartan invariants \cite{MacCallum2015, GANG}. This can be generalized to the concept of an isolated horizon (IH) which arises as a non-expanding horizon (NEH) where a class of null normals, $\{ \ell\}$, exist for which the Lie derivative of $\{\ell\}$ and the induced covariant derivative on the NEH commute \cite{Ashtekar, ashtekar2001, Gourgoulhon2006}. For an IH, the Killing vector field is restricted to the horizon surface and the exterior region may be dynamical. Here, a particular set of SPIs and Cartan invariants vanish on the WIH \cite{ADA, AD} due to the fact that on the horizon the curvature tensor and its covariant derivatives must be of type {\bf II}/{\bf D} relative to the alignment classification \cite{classa,classb,classc}.

It has been conjectured that dynamical black holes admit quasi-local hypersurfaces on which the curvature tensor and its covariant derivatives become more algebraically special. Such a hypersurface, called a {\it geometric horizon} (GH), can be invariantly defined by the vanishing of a particular set of curvature invariants \cite{ADA,AD}. There are examples of dynamical black hole solutions that admit GHs, such as dynamical black hole solutions which are conformally related to a stationary black hole solutions \cite{PageMcNutt,mcnutt2017} and the imploding spherically symmetric metrics \cite{AD}. For the conformal black holes, the event horizon is conformally invariant and can be detected in the original stationary black hole solution. Similarly, for any dynamical spherically symmetric metric the scalar invariant \beq ||\nabla r||^2 = \nabla^a r \nabla_a r,\eeq \noindent where $r$ is the areal radius, will detect the unique, invariantly defined dynamical horizon $r=2M$ \cite{Faraoni:2016xgy}. 

For the spherically symmetric dynamical black holes and dynamical black holes conformally related to stationary black holes the GHs correspond to MTTs. However, in general a GH  will not be a MTT, as the preferred null direction will not necessarily be geodesic and surface forming.  The geometric interpretation of a GH is different from that of a MTT; instead of looking for a spacelike hypersurface constructed from 2D surfaces for which the expansion of the appropriate null normal vector vanishes, we determine the  invariant null coframe adapted to the geometry of the dynamical black hole solution and identify the surfaces where the expansion of the geometrically preferred null vectors vanish which, in turn, affects the algebraic structure of the covariant derivatives of the curvature tensor. It is of interest to determine if less idealized dynamical black hole solutions will admit GHs as well. 

In analogy with the MOTSs and MITSs, we will introduce invariantly defined closed 2D surfaces, called geometrically outer (inner) trapped surfaces (GOTSs or GITSs), for which the expansion scalar $\theta_{(+)}$ ($ \theta_{(-)}$ ) vanishes on spatial hypersurfaces and which make up the GH. While the existence of a GH is not dependent on this foliation, the introduction of GOTSs will be useful for descriptions and illustrations in figures. Similarly, we will say any GH for which $\theta_{(-})<0$ in all GOTSs is a dynamical GH. If a dynamical black hole solution asymptotically evolves to a spherically symmetric dynamical black hole, the dynamical GH will correspond to the dynamical horizon \cite{Faraoni:2016xgy, AD}.

Kastor and Traschen have found a family of exact solutions to the Einstein-Maxwell equations with a cosmological constant representing an arbitrary number of charged $Q=M$ black holes in an otherwise closed universe \cite{KT}. The single mass case corresponds to the Carter black hole solution \cite{carter1973black}. This is the $Q=M$ Reissner-Nordstr{\"o}m-de Sitter solution, which has been studied in \cite{lake1979thin, brill1994global}. In the case of multiple black holes, some aspects of the Kastor-Traschen (KT) solutions have been investigated \cite{brill1994testing}, from which it was concluded that small enough black holes coalesce with each other, while for  mass greater than a critical value there are eternal singularities.

The global structure of the KT solutions has been studied in greater detail and the existence and evolution of marginal surfaces in the case of two equal masses was investigated by Nakao et al \cite{NSH}. The marginal surfaces in these KT solutions belong to four types which bound trapped regions, and hence foliate trapping horizons, which are MTTs with $\theta_{(-)} \neq 0$ and $\mathcal{L}_n \theta_{(+)} \neq 0$ where $n$ is the ingoing null normal \cite{hayward1994general}. The analysis in \cite{NSH} using trapping horizons has implications for the merger and coalescence of multiple black holes. The term ``merger'' denotes the evolution of initially disjoint trapping horizons which become a continuous boundary, while ``coalescence'' denotes the appearance of new marginal surfaces that enclose the original trapped regions \cite{NSH}. If coalescence does not occur, the collision will presumably either produce a naked singularity (violating the cosmic censorship conjecture) or the dynamics will keep the black holes apart.

In principle, the apparent horizons could be used to study the KT solutions, but the analysis would be difficult to implement. If a spacetime admits marginal surfaces, then this is not sufficient to ensure the existence of an apparent horizon \cite{hawking1973large}. The determination of an apparent horizon is problematic since it is necessary to check whether each surface in a given hypersurface is trapped. Furthermore, like the MTTs, the apparent horizon depends on the choice of foliation. While one could determine the connected component of the boundary of an inextendible trapped region, known as the trapping boundary, which would be invariantly defined for a spacetime, in practice this surface is hard to determine numerically \cite{hayward1994general}. 

Motivated by the fact that the event horizons of the Reisner-Nordstr{\"om}-(anti) de Sitter solution are detected by SPIs or Cartan invariants \cite{GANG}, we will investigate the existence of GHs in the multi-black hole four-dimensional (4D) KT solutions using the frame approach and utilizing Cartan invariants. We will compare our results with the results of \cite{NSH} in the case of two black holes, and examine the upper bound on black holes with area larger than $4\pi/\Lambda$ \cite{hayward1994cosmological}. We will also examine the existence of GHs in the three equal mass black hole KT solution to show that these surfaces persist in KT solutions with more than two charged $Q=M$ black holes, and to study the corresponding upper bound on the total mass for such solutions. Finally, we summarize our results and discuss how they provide further evidence for the geometric horizon conjectures \cite{ADA, AD}.

\newpage

\section{The Kastor-Traschen Solution and the Cartan-Karlhede Algorithm} 

The Kastor-Traschen solution represents $N$ charge-equal-to-mass black holes in a spacetime with a positive cosmological constant, $\Lambda$. We will consider the metric in the ``contracting chart'' with $t \in (-\infty, 0)$ \cite{KT,NSH}:  
\begin{eqnarray}
& ds^2=-W^{-2}dt^2+W^2(dx^2+dy^2+dz^2)\,;~W=-Ht+\Sigma_{i=1}^N\frac{m_i}{r_i}. & 
\end{eqnarray}
Here $H=\sqrt{\Lambda/3}$, where $\Lambda\ge0$ is the cosmological constant, $m_i$ ($i\in [1,N]$), are the black hole masses, and $$r_i \equiv \sqrt{(x-x_i)^2+(y-y_i)^2+(z-z_i)^2},$$ are the black hole positions where $r_i = 0$, $i \in[1, N]$, represent a 3D infinite cylinder with 2D cross-sectional area of $4\pi m_i^2$ for each black hole. The electromagnetic 4-potential is given by \beq A=W^{-1}dt. \eeq

For $N>1$, this solution will generically be of Weyl type {\bf I} \cite{ADA, AD}. However,  $F = dA$ gives rise to the following non-zero SPI:

\beq -2 F_{ab} F^{ab} = F^*_{ab} F_*^{ab} = 2 W^{-4} W_{,i} W^{,i},~~i \in [1,3], \eeq

\noindent implying that the electromagnetic field must be non-null, and a coframe exists such that the energy-momentum tensor is of type {\bf D}:

\beq T_{ab} = 4\Phi_1 \bar{\Phi}_1 ( m_{(a} \bar{m}_{b)} + \ell_{(a} n_{b)}). \eeq

\noindent This coframe will be an invariantly defined coframe which can be employed in the Cartan-Karlhede algorithm. To construct this coframe, we start with 

\beq t_0 = \frac{dt}{W},~t_1 = W dx,~t_2 = W dy,~t_3 = W dz, \label{orthonormal} \eeq

\noindent from which we have the null coframe

\beq \begin{aligned} & \ell' = \frac{t_0 - t_1 }{\sqrt{2}},~ n' = \frac{t_0 + t_1 }{\sqrt{2}},~~m' = \frac{t_2 + i t_3}{\sqrt{2}}, \bar{m}' = \frac{t_2 - i t_3}{\sqrt{2}}. \end{aligned} \label{nullframe} \eeq
\noindent Then the electromagnetic field tensor is of the form

\beq F'_{ab} = d(\ell' + n') = d t_0. \eeq

Using the self-dual basis of bivectors ${\bf U}', {\bf V}'$ and ${\bf W}'$ \cite{kramer}: 

\beq \begin{aligned}  & {\bf U}' = 2 {\bf \bar{m}}' \wedge {\bf n}',~~ {\bf V}' =  2 {\bf \ell}' \wedge  {\bf m}' \text{ and } {\bf W}' = 2( {\bf m}' \wedge {\bf \bar{m}}' - {\bf \ell}' \wedge {\bf n}'), \end{aligned} \eeq
\noindent we can express the self-dual electromagnetic field tensor ${F'}^*_{ab}$ as 

\beq \frac12 {F'}^*_{ab} = \Phi'_0 U'_{ab} + \Phi'_1 W'_{ab} + \Phi'_2 V'_{ab}, \eeq

\noindent where 
\beq \begin{aligned} & \Phi'_0 = -\bar{\Phi}'_2 =  -i(\ln W)_{,z} - (\ln W)_{,y} \text{ and } \Phi'_1 = (\ln W)_{,x}. \end{aligned} \eeq

\noindent Applying a null rotation about $n$ and then a null rotation about $\ell$ with their respective parameters defined in \cite{kramer} as,

\beq \begin{aligned} \bar{E} = -\frac{\Phi_1 +\sqrt{\Phi_1^2 + |\Phi_2|^2}}{\bar{\Phi}_0} \text{ and } B = -\frac{\Phi_0}{2(\Phi_1 +\bar{B}\Phi_0}, \end{aligned} \eeq

\noindent produces a new null coframe $\{\ell, n, m, \bar{m}\}$ for which 

\beq F^*_{ab} = 2 \Phi_1 W_{ab} = \frac{\sqrt{W_{,i} W^{,i}}}{W^2} ( m_{[a}\bar{m}_{b]} - \ell_{[a} n_{b]}). \eeq 

\noindent If $N=1$ the Weyl tensor is of Weyl type {\bf D}, and so no further frame fixing is possible at zeroth order. When $N>1$, we may additionally choose a boost and spin so that $\Psi_0 = 1$. While this is necessary for the Cartan-Karlhede algorithm, for our applications we will neglect this choice as it will not affect the form of the Cartan invariants we will use to characterize the GHs. 

The Ricci scalar is $R = 12 H^2$, and relative to this coframe the Ricci tensor takes the form:

\beq R_{ab} =  [4 \Phi_1 \bar{\Phi}_1 + 3 H^2] (m_{(a}\bar{m}_{b)} + \ell_{(a}n_{b)}) = \Phi_{11} (m_{(a}\bar{m}_{b)} + \ell_{(a}n_{b)}). \eeq

\noindent By considering the covariant derivative of the Ricci tensor, two real valued spin-coefficients appear\footnote{These spin-coefficients will also appear in the components of the covariant derivative of the Weyl tensor.}, $\rho$ and $\mu$. From the Bianchi identities, they may be expressed in terms of the components of the Ricci tensor and its covariant derivative:

\beq \rho =  \frac{D \Phi_{11} }{4 \Phi_{11}},~~\mu = \frac{\Delta \Phi_{11}}{4 \Phi_{11}}. \eeq

\noindent These quantities define the expansion of the outcoming and ingoing null vectors of the invariant coframe:

\beq \begin{aligned} & \theta_{(+)} = q^{ab} \ell_{a;b} = Re(\rho + \bar{\rho}) = 2\rho, \\ & \theta_{(-)} = q^{ab} n_{a;b} = -Re(\mu + \bar{\mu}) = -2 \mu, \end{aligned} \eeq

\noindent where $q_{ab} = g_{ab} + 2 \ell_{(a} n_{b)}$ is  the projection operator for $\ell$ and $n$. For spherically symmetric dynamic black holes and black holes admitting NEHs this is also the two-metric induced on the surface $S$ for which $\ell$ and $n$ are normal vectors  \cite{Booth2005}. 

In the case that $N=1$, the coframe can be fixed entirely at first order \cite{GANG}, while in the $N>1$ case, in general, the Weyl tensor is of Weyl type {\bf I}, which is reflected in the non-vanishing of the real SPIs $\mathcal{W}_1$ and $\mathcal{W}_2$ defined in \cite{AD}, in equations (3)-(5). These invariants are constructed from contractions of powers of the Weyl tensor and are equivalent to the real and imaginary parts of the complex invariant $I^3 - 27 J^2$ which is expressed in terms of the complex Weyl tensor in the Newman-Penrose formalism \cite{kramer}. As such the vanishing of $\mathcal{W}_1$ and $\mathcal{W}_2$ is a necessary and sufficient condition for the Weyl tensor  to be of type {\bf II}/{\bf D} \cite{CH, CHDG}. As the coframe can be fully fixed at first order, we will attempt to identify the GHs using the first order Cartan invariants $\rho$ and $\mu$. The hypersurfaces defined by the vanishing of these invariants will be foliation independent.

\section{The Single Mass Kastor-Traschen Solution}

In the case of a single $Q=M$ black hole solution, this is the Reisner-Nordstr{\"o}m-de Sitter black hole. We may choose spherical coordinates for the transverse space, and place the black hole at the origin \cite{brill1994testing}:

\beq ds^2 =  -\frac{dt^2}{W^2}+ W^2 (dr^2 + r^2 d\theta + r^2 \sin^2 \theta d\phi^2),~~W = -H t + \frac{M}{r}. \eeq

\noindent Relative to the null coframe \eqref{nullframe} both the Weyl and Ricci tensor are in the canonical form for type {\bf D} relative to the alignment classification; i.e., the only non-zero Weyl and Ricci spinor components are \cite{kramer}:

\beq \Psi_2 \text{ and } \Phi_{11}. \eeq

\noindent Applying a boost to put the covariant derivative of the Weyl tensor into its canonical form \cite{GANG} relative to these coordinates, the Cartan invariants that detect the horizons are $\theta_{(+)} = \rho \text{ and }\theta_{(-)} = -\mu$ where 

\beq \begin{aligned} \rho &= - \mu = \\ & -\frac{H\sqrt{(H^2 r^2 t^2 +2 HMrt+M^2 - rt)(H^2 r^2 t^2 +2 HMrt+M^2 + rt)}}{\sqrt{2}(Hrt+M)^2}. \end{aligned} \label{N1horizon} \eeq

\noindent The surfaces on which the Cartan invariants $\rho$ and $\mu$ vanish correspond to the surfaces where the timelike Killing vector ${\bf V} = -t \frac{\partial}{\partial t} + r \frac{\partial}{\partial r}$ becomes null, since

\beq |{\bf V}|^2 = \frac{(H^2 r^2 t^2 +2 HMrt+M^2 - rt)(H^2 r^2 t^2 +2 HMrt+M^2 + rt)}{(Hrt+M)^2}. \eeq

\noindent From \eqref{N1horizon}, the GHs related to the outgoing and ingoing null directions coincide. These hypersurfaces correspond to the bifurcate Killing horizons of the Reisner-Nordstr{\"o}m de Sitter black hole \cite{lake1979}, since it is the union of bifurcation surfaces \cite{boyer1969, carrasco2008, chrusciel2012}. It is clear that if $M < \frac{1}{4 H}$, there are three horizons: the inner and outer horizons, and the de Sitter horizon; if $M =  \frac{1}{4H}$, the inner and outer horizons coincide; and if $M >  \frac{1}{4H}$ there is only one horizon.  

\section{The Double Equal Mass Kastor-Traschen Solution} \label{subsec:FW}

The existence of marginal surfaces and trapping horizons has been examined in the case of two coalescing black holes \cite{NSH}. In this case, we can choose coordinates so that the black holes are located on the $x$-axis at a coordinate distance $c>0$ from the origin,
\begin{eqnarray}
r_{\pm}&=&\sqrt{(x\pm c)^2+y^2+z^2}.
\end{eqnarray}

\noindent Due to the upper-bound $4\pi/\Lambda$ on the area of black holes with cosmological constant $\Lambda$, and the fact that the area of a black hole is non-decreasing, this implies that two black holes with total area greater than $ 4\pi/\Lambda$ will not merge, and also imposes a limit on the total mass to be below the critical mass, $M_c = \sqrt{\frac{3}{16\Lambda}} = \frac{1}{4 H}$. It has been shown that if the sum of the two black holes, $$M = m_{+} + m_{-},$$ is below the approximate value, $1.01 M_c$, the black holes will coalesce into a larger single black hole, in the sense that a new future outer trapped horizon appears around the black holes \cite{NSH}. 

For these spacetimes the SPI $\mathcal{W}_2$ vanishes while $\mathcal{W}_1$ is generally non-zero, implying that the Weyl tensor is not globally of Weyl type {\bf II}/{\bf D}. At earliest times, ${\cal W}_1 \to  0$ as $t \to-\infty$, and for finite $t<<0$ there are two 3D GHs enclosing the two black holes which can be located by the vanishing of $\mathcal{W}_1$ \cite{AD}. In the $N=2$ equal mass Kastor-Traschen solution the algebraic type  {\bf II}/{\bf D} discriminant  ${\cal W}_1$ vanishes on segments of the symmetry-axis and at the black hole coordinate locations $r_{\pm}=0$ \cite{AD}. Relative to these coordinates the black holes appear to be points, but they are in fact 2D surfaces of area $4\pi m^2$ for any given time-slice $t = constant$ \cite{NSH}. These denote the horizons of the black holes, and since the flux of matter moving through them is zero, they are isolated horizons. 

We will consider two examples in the contracting chart where $H=0.125$, $m_{\pm} = M/2$ and $c = 0.1$ with $M = 0.5 M_c$ and $M = 1.01 M_c$, to examine the surfaces where the Cartan invariants $\rho$ and $\mu$ vanish. We note that these two examples illustrate the qualitative features of the spacetimes for the subcritical case $M < M_c$ and the supercritical case $M \geq M_c$. Unlike the single mass solution $\rho \neq \mu$ and the explicit form of these Cartan invariants cannot be displayed in a concise form.

\subsection{Subcritical Case: $M = 0.5M_c$} 

We note that the surfaces defined by $\theta_{(-)} =0 $ and the surfaces surrounding the black holes arising from $\theta_{(+)} = 0$ will not intersect for all time-slices. Additionally, within the surfaces defined by $\theta_{(+)} =0 $ the other expansion scalar will be negative; i.e., $\theta_{(-)} < 0 $. The GOTS defined by $\theta_{(+)}=0$ may  not constitute a dynamical GH since $\theta_{(-)} \leq 0$ at isolated points within this surface\footnote{It is possible this is due to numerical error due to Maple and the choice of digits for floating-point numbers.}

\begin{figure}[H] 
  \centering
\begin{subfigure}{0.5\textwidth}
    \includegraphics[scale =0.5]{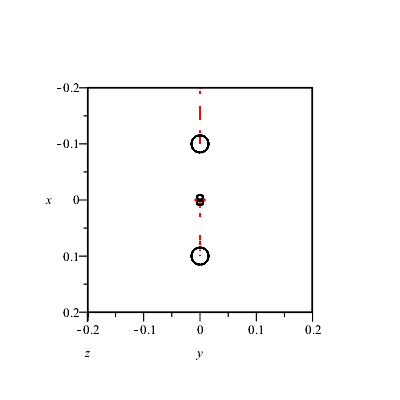}
\end{subfigure} \vspace{-10 mm}
\begin{subfigure}{0.4\textwidth}
    \includegraphics[scale =0.5]{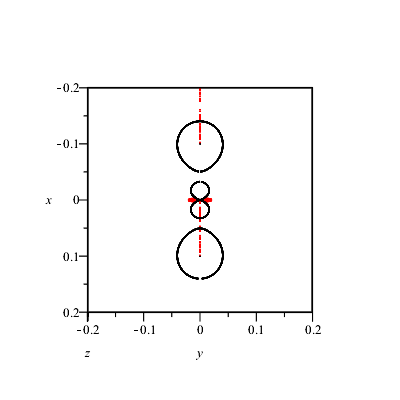}
    \end{subfigure} 
      \caption{Slices of the zeroes of $\rho$ (black) and $-\mu$ (red) at time $t=500t_0$ (left) and $t =180t_0$ (right) with $t_0 = -0.914/H$ in the $z=0$ plane for the $N=2$ subcritical case. } \label{fig:2mslicesub0}
\end{figure}

At early times, the GOTSs located at the coordinate locations of the black holes expand creating spherical GOTSs centred around each of the black holes with additional spherical GOTSs within them. A third GOTS forms around the origin and between the hole of the $\theta_{(-)}=0$ surface which steadily expands, as illustrated in figure \ref{fig:2mslicesub0}.

\begin{figure}[H] 
  \centering
\begin{subfigure}{0.5\textwidth}
    \includegraphics[scale =0.5]{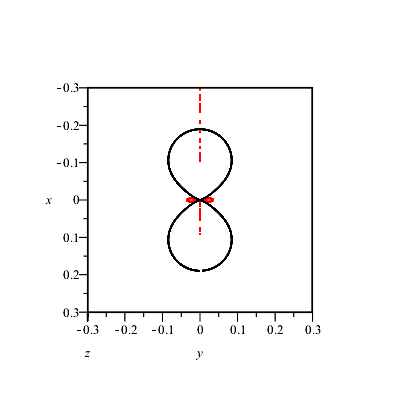}
\end{subfigure} \vspace{-12 mm}
\begin{subfigure}{0.4\textwidth}
    \includegraphics[scale =0.5]{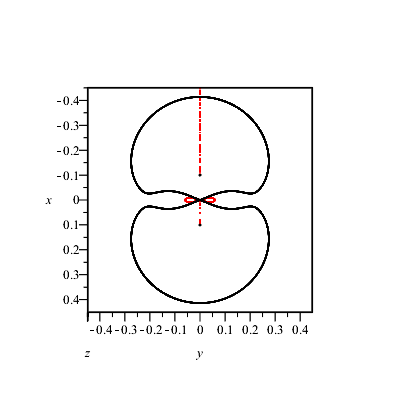}
    \end{subfigure} \vspace{2 mm}
      \caption{Slices of the zeroes of $\rho$ (black) and $-\mu$ (red) at time $t=130t_0$ (left) and $t=70t_0$ (right) with $t_0 = -0.914/H$ in the $z=0$ plane for the $N=2$ subcritical case. } \label{fig:2mslicesub1}
\end{figure}

As time increases, the growing GOTSs combine to make a single  GOTS connected through the hole in the $\theta_{(-)}=0$ surface, while new spherical GOTSs centered on the black hole locations expand. While the black holes move closer together, the outermost GOTS deforms; this is depicted in figure \ref{fig:2mslicesub1} (Note that due to the scale, the GOTSs centered on the black holes are not visible in some of the figures). 
\vspace{-4mm}

\begin{figure}[H] 
  \centering
    \begin{subfigure}{0.5\textwidth}
    \includegraphics[scale =0.5]{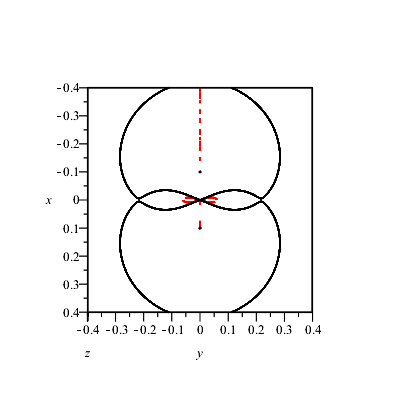}
\end{subfigure} \vspace{-12 mm} \hspace{5 mm}
\begin{subfigure}{0.4\textwidth}
    \includegraphics[scale =0.5]{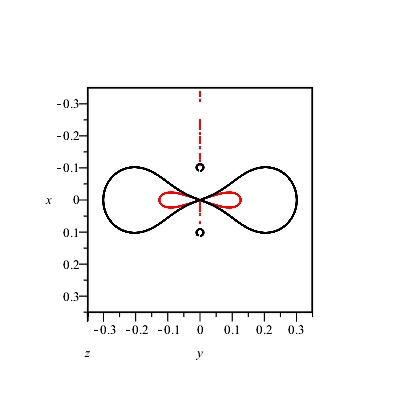}
    \end{subfigure} \vspace{ 2 mm}
      \caption{Slices of the zeroes of $\rho$ (black) and $-\mu$ (red) at time $t= 18.6 t_0$ (left) and $t=t_0 $ (right) with $t_0 = -0.914/H$ in the $z=0$ plane for the $N=2$ subcritical case. } \label{fig:2mslicesub2}
\end{figure}

 \vspace{-3 mm}
As time increases further, the outer GOTS grow outward and deform, forming a torus shaped GOTS aligned along the y-axis, together with a larger GOTS that surrounds all other GOTSs (including the spherical GOTSs centred on each black hole). As time continues, the outermost GOTS expands indefinitely outwards away from the black holes. The start and end of this process is depicted in figure \ref{fig:2mslicesub2}.
 
\begin{figure}[!] 
  \centering 
 \begin{subfigure}{0.5\textwidth} 
    \includegraphics[scale =0.5]{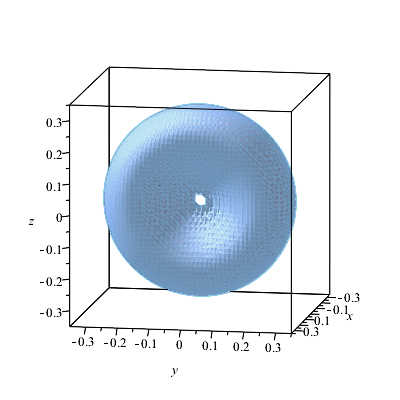}
\end{subfigure}
 \begin{subfigure}{0.4\textwidth} 
  \includegraphics[scale =0.5]{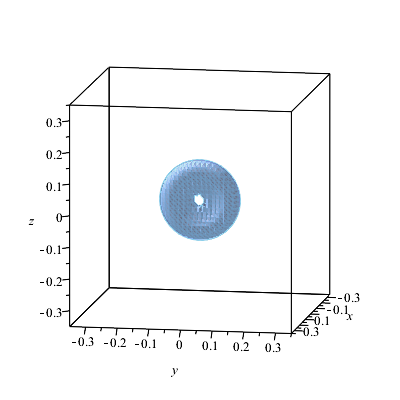}
 \end{subfigure}
  \includegraphics[scale =0.5]{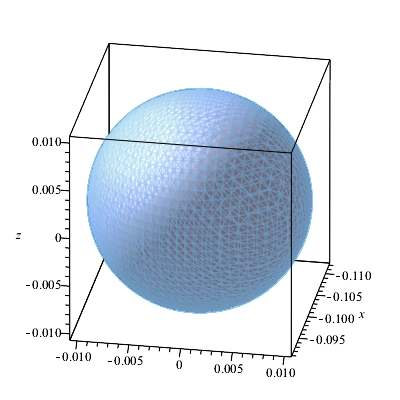}
 \vspace{2 mm}
      \caption{The surfaces defined by the vanishing of $\rho$ (top-left) and $-\mu$ (top-right) in 3D space, at $t=-0.914/H$ for the $N=2$ subcritical case. Due to scale, the surfaces surrounding the black holes are not visible in the graph of $\rho=0$; one of these surfaces defined by $\rho=0$ centered on the black hole location $x=-0.1, y=0, z=0$ is depicted (below). A similar surface is formed around the other black hole location. } \label{fig:2msurf1}
\end{figure}

The choice of time-slices did not allow this process to be shown explicitly during this interval, but the process will repeat a second time.  The 2D surfaces for the time-slice $t=t_0$ defined by $\theta_{(-)}=0$ and $\theta_{(+)}=0$ are depicted in 3D along with one of the smaller surfaces centred on the black holes in figure \ref{fig:2msurf1}, showing that while the outer GOTSs evolve dynamically there are always GOTSs centred on the black holes during this process.

\begin{figure}[H] 
  \centering
\begin{subfigure}{0.5\textwidth}
    \includegraphics[scale =0.5]{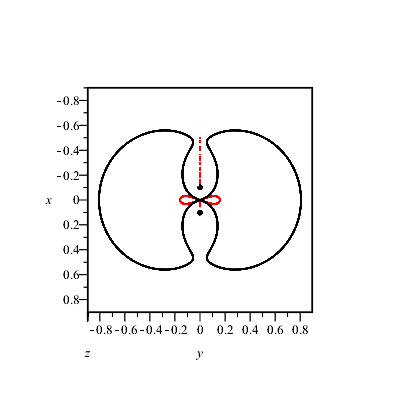}
    \end{subfigure}  
        \begin{subfigure}{0.4\textwidth} 
    \includegraphics[scale =0.5]{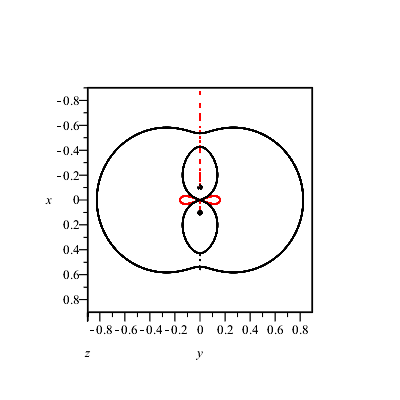}
\end{subfigure} \vspace{-10 mm}
      \caption{Slices of the zeroes of $\rho$ (black) and $-\mu$ (red) at time $t=t_0/4 $ (left), $t= t_0/4.1$ (right) with $t_0 = -0.914/H$ in the $z=0$ plane for the $N=2$ subcritical case. } \label{fig:2mslicesub3}
\end{figure}

The torus-shaped GOTS aligned with the $y$-axis defined by $\theta_{(+)}=0$ now expands outwards and deforms, again forming two additional GOTSs surrounding the spherical GOTSs centred on the black holes and contained within a larger GOTS that expands away from the locations of the black holes. The intermediate GOTSs will merge into one, forming a ``dumb-bell'' shaped GOTS. This is pictured in figure \ref{fig:2mslicesub3}. We note that the merger of the intermediate GOTSs creates a new GOTS surrounding the GOTSs centred on the black holes. 

\begin{figure}[H] 
  \centering
        \begin{subfigure}{0.5\textwidth} 
    \includegraphics[scale =0.5]{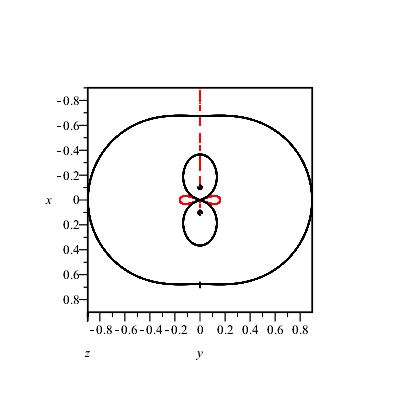}
\end{subfigure}
\begin{subfigure}{0.4\textwidth}
    \includegraphics[scale =0.5]{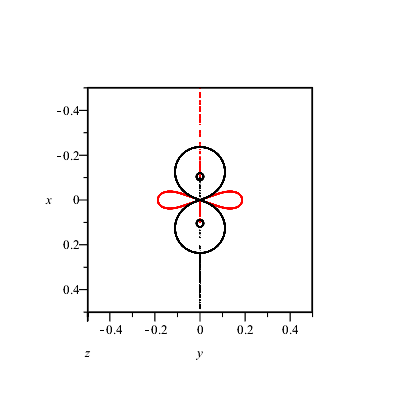}
    \end{subfigure} \vspace{-10 mm}
      \caption{Slices of the zeroes of $\rho$ (black) and $-\mu$ (red) at time $t_0/4.5$ (left) and $t=10^{-10}t_0$ (right) with $t_0 = -0.914/H$ in the $z=0$ plane for the $N=2$ subcritical case. } \label{fig:2mslicesub4}
\end{figure}

\begin{figure}[H] 
  \centering
\begin{subfigure}[h!]{0.45\textwidth}
    \includegraphics[scale =0.45]{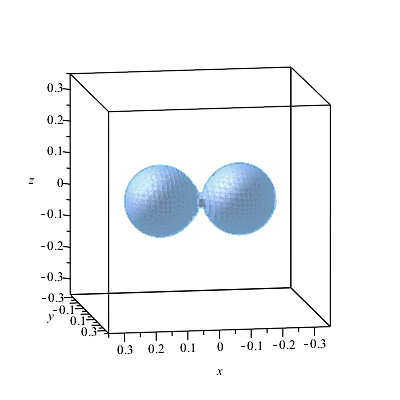}

      \end{subfigure} 
\begin{subfigure}[h!]{0.4\textwidth}
  \centering
    \includegraphics[scale =0.45]{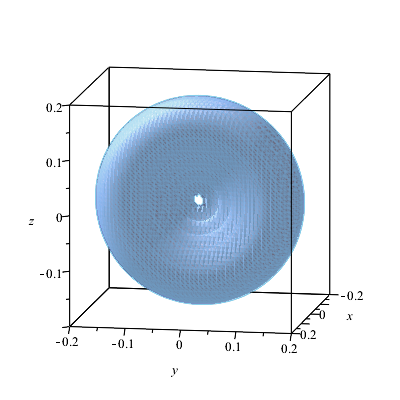}
      \end{subfigure} \vspace{-8 mm}
      \caption{The surfaces defined by the vanishing of $\rho$ (left) and $-\mu$ (right) in 3D space, at $t=10^{-10}t_0$  for the $N=2$ subcritical case.}
      \label{fig:2msurf2}
\end{figure}

When this second process of evolving GOTSs is completed, the innermost GOTSs each constitute dynamical GHs as $\theta_{(-)}<0$, while the outer GOTS make up a GH since $\theta_{(-)}$ is at best non-positive within the surface. This is depicted in figure \ref{fig:2mslicesub4}. The 2D surfaces of this time-slice defined by $\theta_{(-)}=0$ and $\theta_{(+)}=0$ are depicted in 3D in figure \ref{fig:2msurf2}. \vspace{- 3 mm}

After the coalescence of the black holes, the spacetime will eventually settle down to a Reissner-Nordstrom-de Sitter black hole of mass $m_1+m_2$ (which is known to have two GHs \cite{GANG}), since ${\cal W}_1 \to  0$ as  $t \to0^-$ \cite{AD}.

\subsection{Supercritical Case: $M = 1.01 M_c$}

In \cite{NSH} it was shown that two equal mass black holes can combine when the total mass is above the critical mass, $M_c$. In particular, it was shown that for $M = 1.01 M_c$ the  black holes will coalesce and that at late times the outer marginal surface vanishes. Due to the expectation that KT multi-black hole solutions with total mass $M \geq M_c$ correspond to spacetimes with naked singularities and hence should not be able to merge at any time due to the upper-bound on the area of the resulting single black hole, this suggests that in the supercritical case the black holes can potentially coalesce as a new marginal surface temporarily forms around them.

\begin{figure}[H]  
  \centering
\begin{subfigure}{0.5\textwidth}
    \includegraphics[scale =0.49]{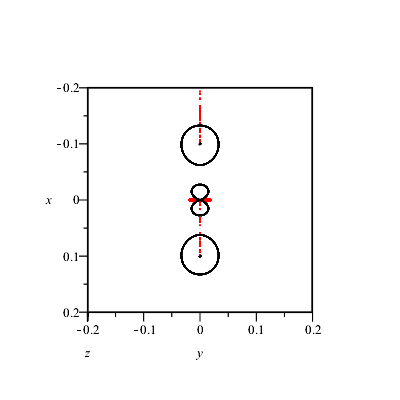}
\end{subfigure} 
\begin{subfigure}{0.4\textwidth}
    \includegraphics[scale =0.49]{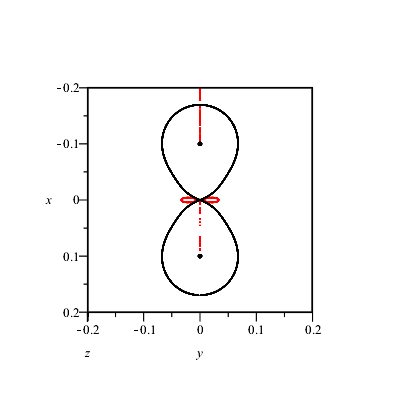}
    \end{subfigure}   \vspace{-10 mm}
      \caption{Slices of the zeroes of $\rho$ (black) and $-\mu$ (red) at time $t=180t_0$ (left) and $t=70t_0$ (right) with $t_0 = -0.914/H$ in the $z=0$ plane for the $N=2$ supercritical case. } \label{fig:2mslicesuper1}
\end{figure}

\begin{figure}[H]  
  \centering
    \begin{subfigure}{0.5\textwidth}
    \includegraphics[scale =0.5]{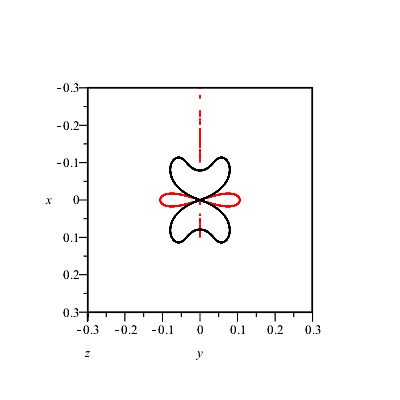}
\end{subfigure} 
\begin{subfigure}{0.4\textwidth}
    \includegraphics[scale =0.5]{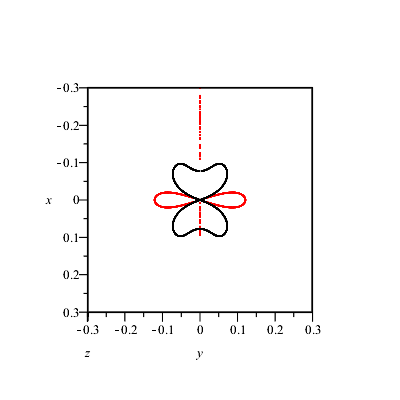}
    \end{subfigure}   \vspace{-10 mm}
      \caption{Slices of the zeroes of $\rho$ (black) and $-\mu$ (red) at time $t=t_0 = -0.914/H$ (left), $t = 10^{-10}t_0$ (right)   in the $z=0$ plane for the $N=2$ supercritical case. } \label{fig:2mslicesuper2}
\end{figure}

At early times the GOTSs in the supercritical case will behave in a similar manner to that in the subcritical case, yielding spherical GOTSs centred on the black holes contained within a larger GOTS; this is illustrated in figure \ref{fig:2mslicesuper1}.
\vspace{-3 mm}
However, around $t = 50t_0$ the evolution of the outermost GOTS differs significantly. Unlike the MTT in the supercritical case, that appears when the black holes coalesce and vanishes at late times \cite{NSH}, the outermost GH that forms around the other GHs does not vanish. Instead, this surface deforms by contracting along the $x$-axis towards the origin and exposes the black holes. The inner GOTSs centred around the black hole masses which exist at early times, no longer exist after the outer GOTS has been pulled back. The end state of this process is depicted in figure \ref{fig:2mslicesuper2}.

\section{The Triple Equal Mass Kastor-Traschen Solution}

For $N=3$ we will consider three black holes with equivalent masses and critical mass $M_c = 4$, which gives the corresponding value $H = 1/16$. In this case, two of the black holes are placed on the $x$-axis, and the third lies on the $y$-axis in the following manner:
\beq & r_{\pm} = \sqrt{(x \pm c)^2+y^2+z^2},~~ r_{3} = \sqrt{x^2+(y - c)^2+z^2}. & \eeq

For illustration, we will consider two examples in the contracting chart where, $m_i = M/3,~i \in[1,3]$ and $c = 0.1$ with $M = 0.75 M_c$ and $M = 1.5 M_c$. As in the $N=1$ and $N=2$ cases we employ the invariant coframe determined by the Cartan-Karlhede algorithm, and examine where the extended Cartan invariants $\rho$ and $\mu$ vanish at fixed time slices. We note that these two examples illustrate the qualitative features of the spacetimes for the subcritical case $M < 1.5 M_c$ and the supercritical case $M \geq 1.5 M_c$.

\subsection{Subcritical Case, $M = 0.75M_c$} 

\begin{figure}[H] 
  \centering
 \begin{subfigure}[h!]{0.5\textwidth}
    \includegraphics[scale =0.5]{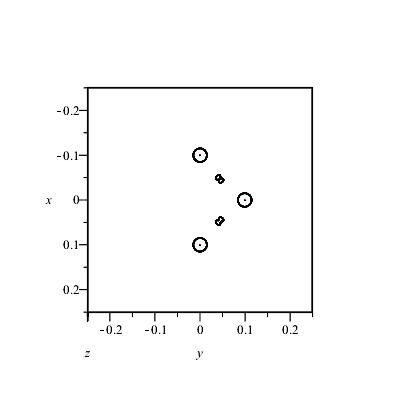}
 \end{subfigure} 
 \begin{subfigure}[h!]{0.4\textwidth}
    \includegraphics[scale =0.5]{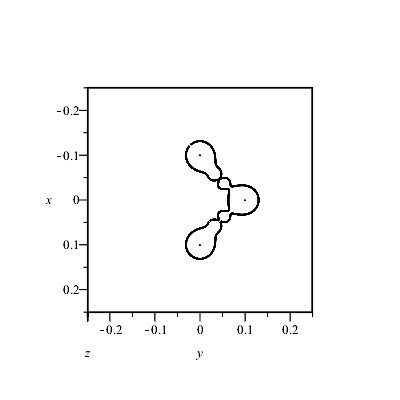}
 \end{subfigure}  \vspace{-10 mm}
 \caption{Slices of the zeroes of $\rho$ (black) and $-\mu$ (red) at time $t=1000t_0$ (left), $t=500t_0$ (right) with $t_0 = -0.914/H$ in the $z=0$ plane for the $N=3$ subcritical case. } \label{fig:3mslicesub0}
\end{figure}

The evolution of the GOTSs in the subcritical case of three black holes is similar to the case of two black holes. At early times, around each black hole an inner spherical GOTS forms, while an outer spherical GOTS gradually grows larger for each black hole. Along the two lines with equal length of the isosceles triangle formed by the black holes' locations, a GOTS forms at the center-point of each line.  As the black holes near, the outer spherical GOTSs merge with the expanding GOTSs lying on the equal length lines of the isosceles triangle and forms a single outermost GOTS. This is shown in figure \ref{fig:3mslicesub0}.

\begin{figure}[H] 
  \centering
 \begin{subfigure}[h!]{0.5\textwidth}
    \includegraphics[scale =0.5]{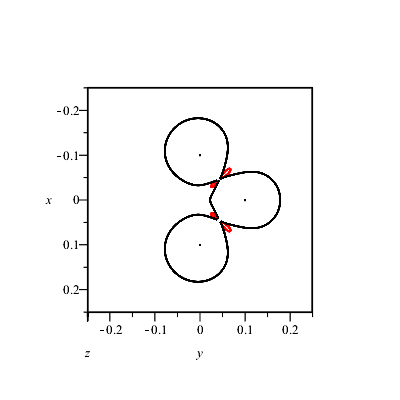}
 \end{subfigure} 
 \begin{subfigure}[h!]{0.4\textwidth}
    \includegraphics[scale =0.5]{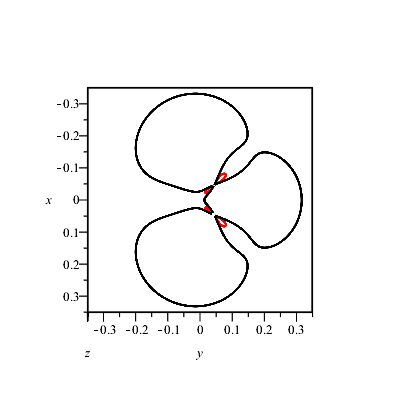}
 \end{subfigure}  \vspace{-10 mm}
 \caption{Slices of the zeroes of $\rho$ (black) and $-\mu$ (red) at time $t=130t_0$ (left), $t=40t_0$ (right) with $t_0 = -0.914/H$ in the $z=0$ plane for the $N=3$ subcritical case. } \label{fig:3mslicesub1}
\end{figure}

\begin{figure}[H] 
  \centering
 \begin{subfigure}{0.5\textwidth}
    \includegraphics[scale =0.5]{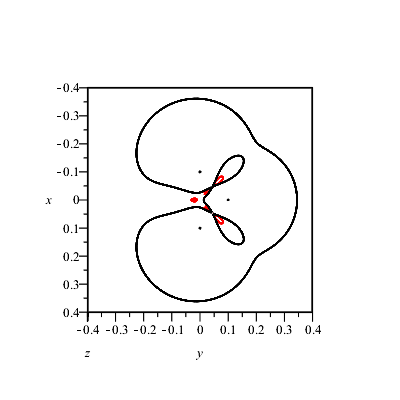}
 \end{subfigure} 
 \begin{subfigure}{0.4\textwidth}
    \includegraphics[scale =0.5]{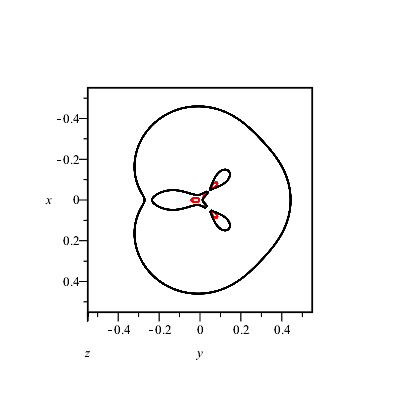}
 \end{subfigure}  \vspace{-10 mm}
 \caption{Slices of the zeroes of $\rho$ (black) and $-\mu$ (red) at time $t=35t_0$ (left), $t=25 t_0 $ (right) with $t_0 = -0.914/H$ in the $z=0$ plane  for the $N=3$ subcritical case. } \label{fig:3mslicesub2}
\end{figure}

This newly formed outermost GOTS initially does not intersect the surface defined by $\theta_{(-)}=0$, until it begins to expand outwards and deform. This is depicted in figures \ref{fig:3mslicesub1} and \ref{fig:3mslicesub2}.

As this outermost GOTS expands, it deforms into a surface with spherical topology that then expands outwards to spatial infinity. During this process, the spherical GOTSs centred on each black hole remains. This is depicted in figure \ref{fig:3mslicesub2}

\begin{figure}[H] 
  \centering
 \begin{subfigure}{0.5\textwidth}
    \includegraphics[scale =0.5]{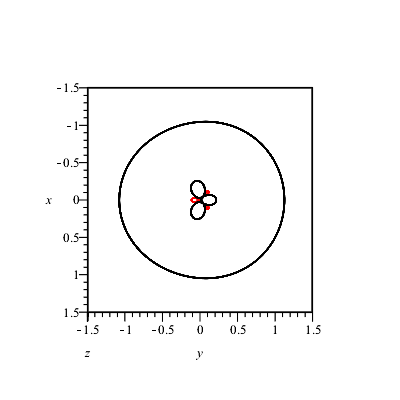}
 \end{subfigure} 
 \begin{subfigure}{0.4\textwidth}
    \includegraphics[scale =0.5]{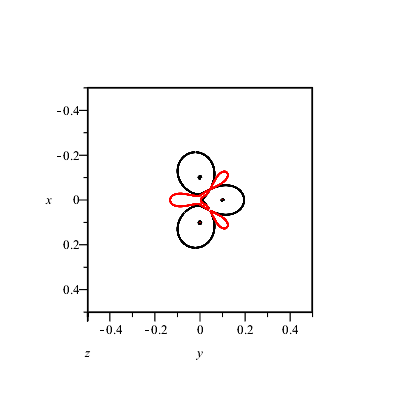}
 \end{subfigure}   \vspace{-10 mm}
 \caption{Slices of the zeroes of $\rho$ (black) and $-\mu$ (red) at time $t=t_0$ (left) and $t=10^{-10}t_0$ (right)  with $t_0 = -0.914/H$ in the $z=0$ plane. } \label{fig:3mslicesub3}
\end{figure}

While the outermost GOTS expands outwards, the inner GOTSs expand to form a connected surface, and within this new connected surface new spherical GOTSs form around each black hole. We note that this connected surface will not deform, but instead will remain a connected region that does not intersect with the surface defined by $\theta_{(-)} =0$, as shown in figure \ref{fig:3mslicesub3}

\begin{figure}[H]  
  \centering
  \begin{subfigure}[h!]{0.5\textwidth}
    \includegraphics[scale =0.45]{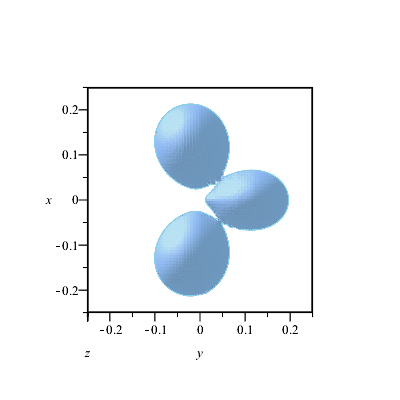}
  \end{subfigure} 
  \begin{subfigure}[h!]{0.4\textwidth}
        \includegraphics[scale =0.45]{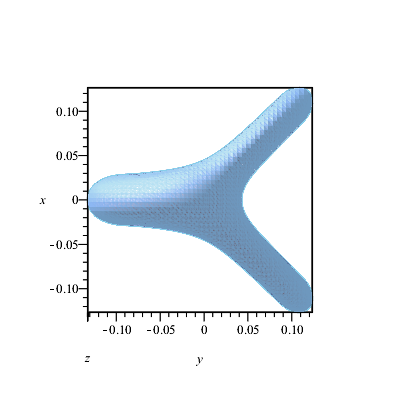}
 \end{subfigure}  \vspace{-8 mm}
      \caption{The surfaces surrounding the 3 black holes defined by the vanishing of $\rho$ (left) and $-\mu$ (right)   at $t_0=-0.914/H$ viewed from above.} \label{fig:3msurftopview}
\end{figure}

\begin{figure}[H] 
  \centering
  \begin{subfigure}[h!]{0.5\textwidth}
    \includegraphics[scale =0.45]{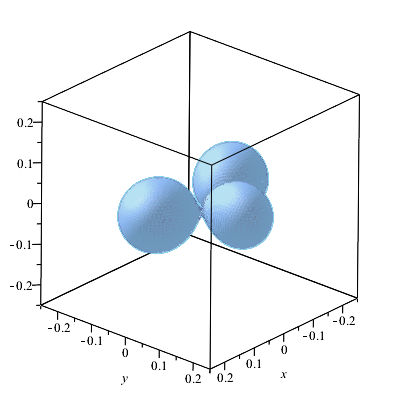}
    \end{subfigure}  
    \begin{subfigure}[h!]{0.4\textwidth}
    
    \includegraphics[scale =0.45]{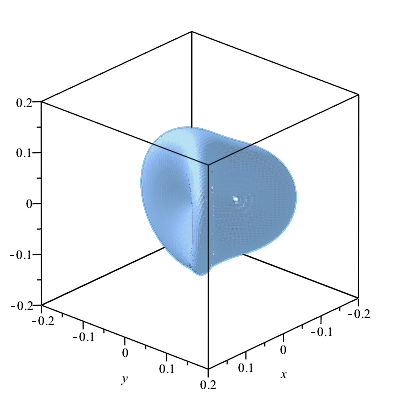}

\end{subfigure}   

      \caption{The surfaces surrounding the 3 black holes defined by the vanishing of $\rho$ (left) and  $-\mu$ (right) in 3D space, at $t = 10^{-10}t_0$.} \label{fig:3msurfrhomu}
\end{figure}

At late times there are GOTSs centred around each black hole, and an outer connected GOTS surrounding them, which does not intersect the surface defined by $\theta_{(-)} = 0$. These GOTSs together form dynamical GHs as they do not intersect with the surfaces defined by  $\theta_{(-)}=0$ and therefore must have $\theta_{(-)}<0$ within them. The outermost surfaces defined by $\theta_{(-)}=0$ and $\theta_{(+)}=0$, respectively, at $t_0 = -0.914/H$ and $t= 10^{-10} t_0$ are displayed in 3D in figures \ref{fig:3msurftopview} and \ref{fig:3msurfrhomu} to illustrate the fixed nature of these surfaces at late times. 

\subsection{Supercritical Case $M = 1.5 M_c$}

Following the work of \cite{NSH,brill1994testing}, there was an expectation that the GHs would behave in a similar manner to the $N=2$ case. Surprisingly, in the case of three black holes in the KT solutions the behaviour is different at late times. Instead of a strict inequality $M < M_c$, the total mass of the three black holes may exceed the critical mass, giving the upper bound $M \leq 1.5 M_c$.

\begin{figure}[H] 
  \centering
\begin{subfigure}[h!]{0.5\textwidth}
    \includegraphics[scale =0.5]{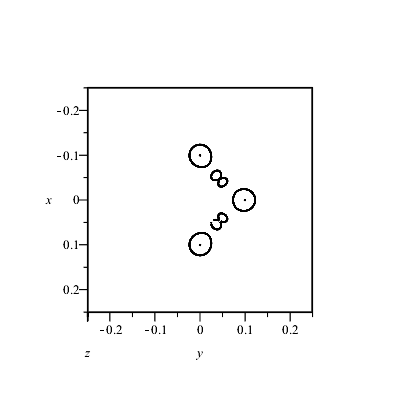}
\end{subfigure} 
\begin{subfigure}[h!]{0.4\textwidth}
    \includegraphics[scale =0.5]{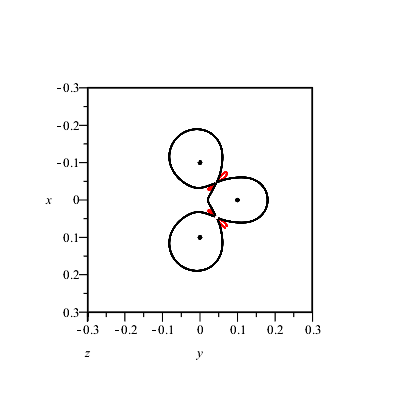}
\end{subfigure}  \vspace{-10 mm}
      \caption{Slices of the zeroes of $\rho$ (black) and $-\mu$ (red) at time $t=500t_0$ (left), $t=70t_0$ (right) with $t_0 = -0.914/H$ in the $z=0$ plane for the $N=3$ supercritical case. } \label{fig:3mslicesuper1}
\end{figure}

\begin{figure}[H] 
  \centering
\begin{subfigure}{0.5\textwidth}
    \includegraphics[scale =0.5]{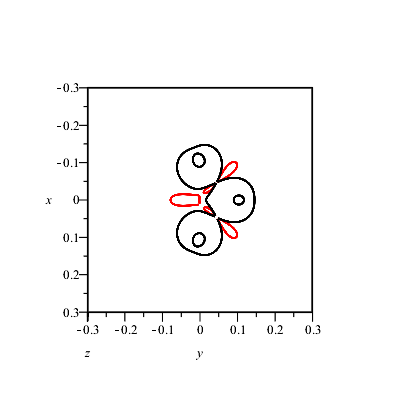}
\end{subfigure}
\begin{subfigure}{0.4\textwidth}
    \includegraphics[scale =0.5]{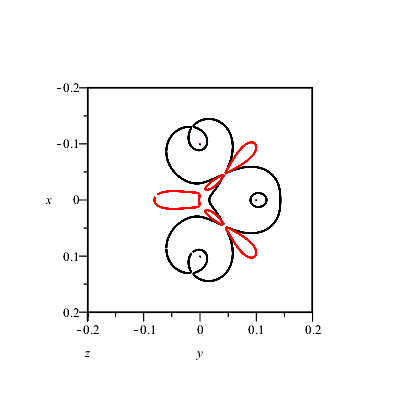}
\end{subfigure}   \vspace{-10 mm}
      \caption{Slices of the zeroes of $\rho$ (black) and $-\mu$ (red) at time $t=0.25 t_0$ (left) and $t=10^{-10}t_0$ (right)  with $t_0 = -0.914/H$ in the $z=0$ plane for the $N=3$ supercritical case. } \label{fig:3mslicesuper2}
\end{figure}

As in the $N=2$ case, the behaviour of the surfaces is similar to the subcritical case until late times when certain parts of the connected outer GOTS pull back, exposing two of the black holes. Unlike the $N=2$ case, the exposed black holes maintain a GOTS centred around each of them.  This behaviour is depicted in figures \ref{fig:3mslicesuper1} and \ref{fig:3mslicesuper2}. This appears to be generic behaviour for $M_c \leq M \leq 1.5 M_c$. While for $M > 1.5M_c$, the black holes coalesce temporarily but the outer GOTS pulls back and removes the inner GOTSs leaving the black holes exposed.

\newpage 

\section{Summary and Discussion} 

In this paper we have utilized the necessary steps of the Cartan-Karlhede algorithm needed to construct an invariant coframe and generated curvature invariants that can be used to determine the geometrical properties of the Kastor-Traschen multi-black hole solutions. Using the invariant null coframe, we have shown that the expansion scalars, $\theta_{(+)}$ and $\theta_{(-)}$, of the geometrically preferred outgoing and ingoing null vectors  $\ell$ and $n$, are extended Cartan invariants. Due to their geometrical interpretation and their appearance in the covariant derivatives of the curvature tensor, we examined where these curvature invariants vanish and found hypersurfaces which are necessarily foliation independent, implying that they are GHs. The existence of GHs bounding the black holes in each of the examples gives further support to the geometric horizon conjectures \cite{ADA, AD}. 

In general, GHs will not be apparent horizons, dynamical horizons or MTTs. However, for spherically symmetric dynamical black hole solutions GHs will coincide with the unique dynamical horizon $r=2M$ \cite{Faraoni:2016xgy, AD}. Furthermore, if a dynamical black hole solution settles down to a dynamical solution where the black hole is no longer interacting with the exterior region, then the GHs will coincide with IHs  \cite{booth2007isolated, AshtekarKrishnan}. For example, in the subcritical Kastor-Traschen solutions which do not contain naked singularities, after the $N$ black holes have merged the spacetime will eventually settle down to a type {\bf D} Reissner-Nordstr{\"o}m-de Sitter black hole with mass $M=\Sigma_{i} m_i$, implying that in the quasi-stationary regime there will be a single GH. This suggests that by tracking the GHs that arise in a dynamical black hole solution we can employ one of these hypersurfaces to determine a smooth, dynamical hypersurface that shields all other horizons and identifies the region of interest \cite{ADA, AD}.

The vanishing of $\theta_{(+)} $ and $\theta_{(-)} $ provide several distinct hypersurfaces that evolve over time. By studying the sign difference on either side of the surfaces defined by $\theta_{(\pm)} =0$ it is possible to determine if and when a given hypersurface is a dynamical GH, which would be expected to evolve into a dynamical horizon or an IH at later times \cite{Booth2005,Senov}. The subcritical examples for the $N=2$ and $N=3$ cases show that the outermost GH  may not be a global dynamical GH due to the possibility that $\theta_{(-)} \leq 0$ (instead of a strict inequality), but this needs further numerical confirmation.  

The GHs that surround the black holes evolve as would be expected for collapsing black holes. In particular, in the $N=2$ case, the upper-bound on the total mass, $M$, is strict, where for $M \geq M_c$ the GH eventually moves away from the black holes, potentially leaving naked singularities. In the $N=3$ case, we have found that for a total mass $M > 1.5 M_c$ an outer GH forms around the black holes, but eventually recedes leaving some of the black holes exposed, without any GHs around them.

The goal of the present analysis is to provide motivation for the use of GHs in the KT solutions by demonstrating that they behave in a similar manner to the known foliation-dependent quasi-local horizons with regards to the upper-bound on the total mass \cite{brill1994testing,NSH}. In addition, we have shown that by choosing an invariant coframe constructed from the principal null directions of the curvature tensor, the vanishing of the Cartan invariants $\rho$ and $\mu$ on the GHs affect the form of the covariant derivatives of the curvature tensor in accordance with the geometric horizon conjectures \cite{ADA, AD}.

In future work we will examine the surface area, intrinsic curvature and extrinsic curvature of the surfaces as they evolve in time in order to study the dynamics of the GHs surrounding the black holes. We will also explore the relevance of other extended Cartan invariants in order to determine analogues for the scalars related to the flux of energy across the dynamical horizons  \cite{AshtekarKrishnan,AK1,AK2} which correspond to the NP coefficients $\sigma$ and $\lambda$ relative to the coframe adapted to the null normal vector fields of the dynamical horizon. Assuming there are indeed Cartan invariants that describe the flux of energy across the GHs, it is of interest to track their evolution in order to study the rate of area increase for the GHs in a similar manner to event horizons \cite{hawking1972energy} and dynamical horizons  \cite{booth2007isolated, gupta2018dynamics}.

These issues will also be explored for other dynamical spacetimes such as, for example, the quasi-spherical Szekeres solutions \cite{gaspar2018black}, the extreme-mass ratio limit of a binary black hole merger \cite{Emparan2016,hussain2017deformation,emparan2018black} and dynamical solutions conformally related to static multi-black hole solutions, such as the Majumdar-Papapetrou (MP) solutions \cite{papapetrou1945static, majumdar1947class,gurses1998sources}. In order to investigate the GH conjectures for  numerical solutions with analytic initial data \cite{nakao1993apparent}, or numerical black hole solutions, we must either consider an implementation of a covariant frame formalism for numerical relativity  \cite{hamilton2017covariant} or investigate the possibility that the curvature invariants dictating the expansion of the outgoing and ingoing null vectors can be expressed in terms of SPIs. Since spacetimes are $\mathcal{I}$-non-degenerate they are locally characterized by both their Cartan invariants and the set of SPIs \cite{CSI4a} and it is expected that it is indeed possible to express the Cartan invariants in terms of SPIs \cite{CSI4a,OP}.

\section*{Acknowledgements} 

We would like to thank Andrey Shoom for useful discussions at the beginning of this project. The work was supported by NSERC of Canada (A.C.), and through the Research Council of Norway, Toppforsk grant no. 250367: Pseudo- Riemannian Geometry and Polynomial Curvature Invariants: Classification, Characterisation and Applications (D.M.).

\bibliographystyle{unsrt-phys}
\bibliography{GHReferences}

\end{document}